\documentstyle[epsf,epsfig,12pt]{article}
\begin{document}
\pagestyle{empty}
\begin{flushright}
{SHEP-98/06}\\
\end{flushright}
\vspace*{5mm}
\begin{center}
{\large {\bf The Effect of Higher Order Corrections to the BFKL Equation
 on the Perturbative Pomeron}}\\
\vspace*{1cm}
{\bf D. A. Ross} \\
\vspace{0.3cm}
Physics Department,\\
University of Southampton,\\
Southampton SO17 1BJ, United Kingdom\\
\vspace*{2cm}
{\bf ABSTRACT} \\ \end{center}
\vspace*{5mm}
\noindent
We discuss the significance of the next-to-leading order term in the BFKL
equation on the energy dependence of diffractive processes controlled
by the perturbative QCD pomeron. It is shown that whereas the large negative
corrections do indeed reduce the rate of growth of diffractive amplitudes
with increasing energy, this reduction is considerably less than previously
expected. 

\vspace*{5cm}

\begin{flushleft}
SHEP-98/06 \\
April 1998
\end{flushleft}
\vfill\eject

\setcounter{page}{1}
\pagestyle{plain}

In a recent paper Fadin and Lipatov \cite{FL}  have presented,
 in compact form, the
results of the higher order corrections to the BFKL \cite{bfkl,BL} equation.
The solution to this equation is the amplitude,
$f(s,t,{\mathbf k, \, k^\prime})$, for the scattering of two gluons
with transverse momenta ${\mathbf k}$ and ${\mathbf k^\prime}$ respectively,
 centre-of-mass energy $\sqrt{s}$ and square momentum transfer $t$,
in the `diffractive region' where $ s \gg |t|$. It may be written
\begin{equation}
\frac{\partial}{\partial \ln s} f(s,t,{\mathbf k, \, k^\prime})
= \int d^2{\mathbf l} \,  {\cal  K}(t,{\mathbf k, \, l})
f(s,t,\mathbf{l, \, k^\prime}). \label{eq1} \end{equation}
We are interested here in the forward scattering amplitude required
for total cross-sections or deep-inelastic structure functions and so we
henceforth set $t=0$ and suppress it.

The solutions to Eq.(\ref{eq1}) are found through the identification
of a complete set of eigenfunctions, $\phi_\nu(\mathbf{k})$,
of the kernel ${\cal K}$ and their corresponding eigenvalues, $\omega(\nu)$
\begin{equation}
\int d^2{\mathbf l}  \, {\cal  K}({\mathbf k,l})
\phi_\nu({\mathbf l})
= \omega(\nu) \phi_\nu({\mathbf l})
, \label{eq2} \end{equation}
where,   up to order $\alpha_s^2$,
\begin{equation}
\omega(\nu)=\bar{\alpha} \chi^{(1)}(\nu) + \bar{\alpha}^2 \chi^{(2)}(\nu),
\label{eq3} \end{equation}
with $\bar{\alpha}=3\alpha_s/\pi$.
The  eigenfunctions  $\phi_\nu({\mathbf k})$ are given by
\begin{equation}
    \phi_\nu({\mathbf k})= \frac{1}{\sqrt{ {\mathbf k}^2}}
     \left(\frac{{\mathbf k}^2}{\sqrt{\alpha_s({\mathbf k}^2)}} 
 \right)^{i \nu}
     \label{eq4} \end{equation}
 where we have taken the advice in ref. \cite{FL} and included in $\phi_\nu$
 a factor which breaks the conformal invariance in  such a way that
 it has no effect on the leading order eigenvalue but guarantees
 an eigenvalue  to next order which is a (real) even function of $\nu$.

It was pointed out in ref. \cite{FL} that the next order correction
is large and of opposite sign, leading to a considerable reduction in the
high energy (low-$x$) dependence of processes mediated by this amplitude
( the  `perturbative pomeron').
Indeed if we look at $\nu=0$ we find
\footnote{The second  order term in $\chi$  has a mild dependence on the
number of active flavours. We set this number to three.}
\begin{equation}
\chi(0)=2 .77 \bar{\alpha} - 18.34 \bar{\alpha}^2, \label{eq5} \end{equation}
which changes sign for the small value of 0.16 for $\alpha$. The immediate
conclusion is that even at HERA,
 where the running coupling is sufficiently small
for a perturbative expansion to be reliable, the first two terms in the
calculation of $\chi$ are insufficient.
This may well be the case. However, in the absence (presumably indefinitely)
of further terms in the expansion of $\chi$, 
one has no option but to assume that the expansion calculated
so far represents a valid approximation and
it is interesting
to examine in more detail the consequences of the next order
corrections for the $s$ dependence of diffractive amplitudes.

\begin{figure}
\begin{center}
\leavevmode
\hbox{\epsfxsize=4.8 in
\epsfbox{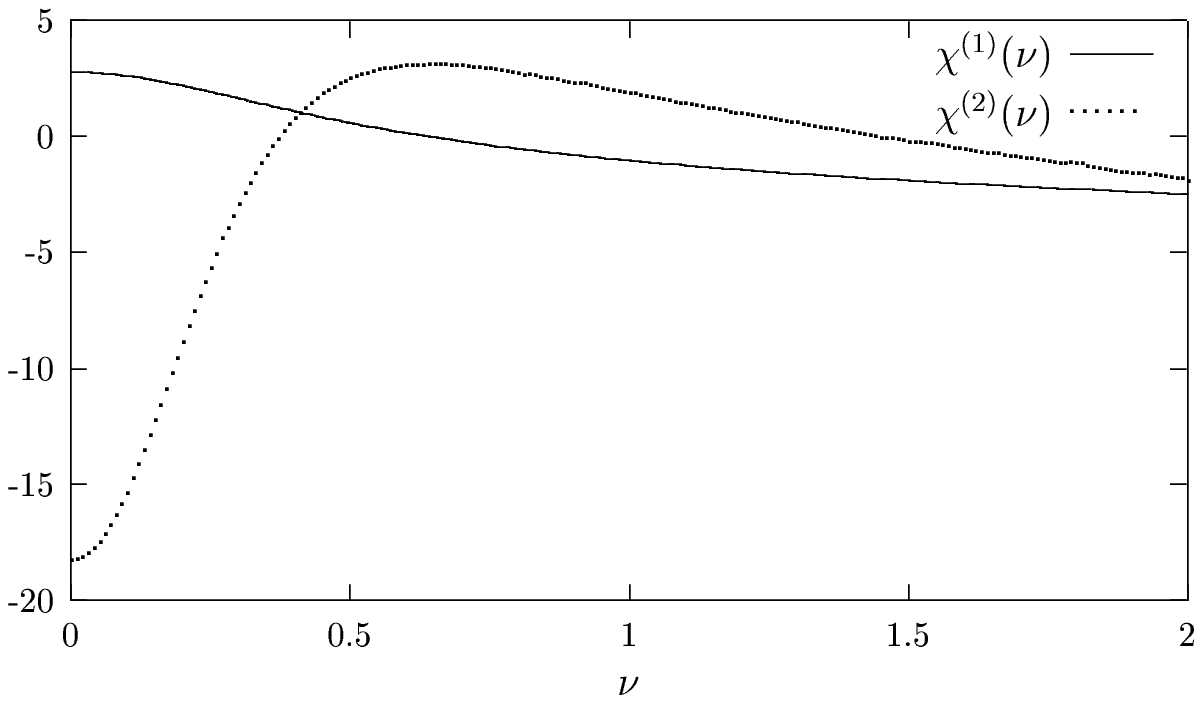}}
\caption{}
\end{center}
\end{figure}

In Fig. 1 we have plotted
the functions $\chi^{(1)}$ (solid line) and $\chi^{(2)}$ (dotted line)
against $\nu$ in the region $0 < \nu < 2$. We observe that the higher order
correction is largest for $\nu=0$ but that its magnitude diminishes rapidly
as $\nu$ increases. Moreover, unlike $\chi^{(1)}$, $\chi^{(2)}$ has a turning
point at $\nu \approx 0.6$. This is consistent with the asymptotic values, 
which are of the same sign, namely
$$ \lim_{\nu \to \infty} \chi^{(1)}(\nu) = -2 \ln \nu $$
$$ \lim_{\nu \to \infty} \chi^{(2)}(\nu) = - \left( \frac{22}{3}-
 4\frac{n_f}{9} \right) \ln^2(\nu) $$ 

Fig.1 shows us that it is {\it not} the $\nu=0$ component of the function
$f(s,{\mathbf k, \, k^\prime})$ that dominates (except for exceedingly small
values of $\alpha_s$). Indeed we see that the second order term has a
positive  maximum for $\nu \approx 0.6$, so that as $\alpha_s$ increases
the intercept of the QCD pomeron also increases. There is no maximum 
value as there is for the $\nu=0$ component.

\begin{figure}
\begin{center}
\leavevmode
\hbox{\epsfxsize=4.8 in
\epsfbox{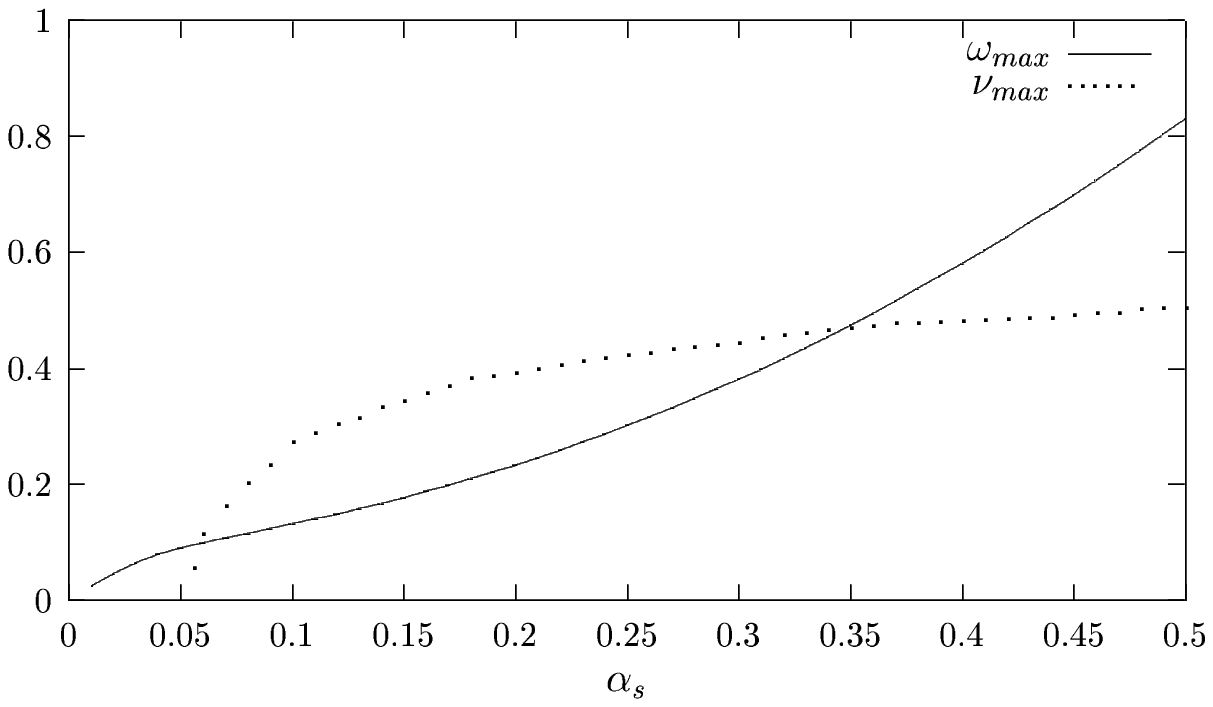}}
\caption{}
\end{center}
\end{figure}

 The maximum value of $\omega(\nu)$, which
 determines the high energy behaviour of the pomeron amplitude occurs
at a value of $\nu$ which depends on the coupling constant $\alpha_s$.
In Fig. 2 we have plotted both this maximum value, $\omega_{\max}$
and the value $\nu_{max}$ at which this occurs for values of $\alpha_s$
up to 0.5, which is the largest value for which one might reasonably
expect to be able to rely on a perturbative calculation.
The increasing value of $\omega_{\max}$ with $\alpha_s$ is unfortunate from
the point of view of those who might have hoped that the maximum
value of  0.1 for $\omega(0)$, which is very little 
greater than the intercept of the `soft' pomeron \cite{DL}, might have contained a clue to the reconciliation of the soft and hard pomerons from the higher
 order corrections. On the other hand it must be noted that since the
larger values of $\omega_{max}$ refer to larger values of $\nu$, the components
which dominate at large $s$ will have an oscillatory behaviour in
transverse momenta and will therefore be somewhat suppressed when the
pomeron amplitude is convoluted with the impact factors that determine
the coupling of the pomeron to the scattering hadrons \cite{BL}. The upshot
of this is that one may have to go to extremely large energies before
these large $\omega$ components dominate. Such a scenario was proposed
in ref.\cite{cdl}. 

More importantly, the large $s$ behaviour of the perturbative pomeron
is not suppressed by the higher order corrections as much as has initially
been anticipated. The steep change of $\chi^{(2)}(\nu)$ near $\nu=0$
shown in Fig. 1 means that for all but extremely small values of $\alpha_s$,
an expansion of $\omega$ up to quadratic order in $\nu$ followed by a saddle
point integration is not an appropriate technique for finding the solution
to Eq.(\ref{eq1}) up to ${\cal O}(1/\ln^2s)$. Instead we must expand up
to fourth order in $\nu$ since it is this fourth order term that is negative.
We find
\begin{equation}
\chi^{(1)}(\nu)= 4 \ln 2 - 14 \zeta(3) \nu^2 \ +  \ 62 \zeta(5) \nu^4 \
 + \ \cdots , 
\label{eq6} \end{equation}
where
$$ \zeta(n) \ = \ \sum_{i=1}^\infty \frac{1}{i^n}, $$
is the Riemann zeta function, and
\begin{eqnarray}
\chi^{(2)}(\nu)&=& -h_0-\frac{27}{128}\pi^3-\frac{11}{2} \zeta(3)
+\frac{67}{9} \ln(2) -\frac{1}{3}\pi^2 \ln(2) -\frac{22}{3} \ln^2(2)  \nonumber
 \\ & & \hspace*{1cm}  
-n_f \left[ \frac{11}{3456}\pi^3 +\frac{10}{27} \ln (2) -\frac{4}{9} \ln^2 (2)
 \right] \nonumber \\ & & 
 +\left( -h_2+\frac{199}{768} \pi^5 -\frac{1}{128} \pi^3 +186 \zeta(5)
 -\frac{469}{18} \zeta(3) +\frac{7}{6} \pi^2 +\frac{154}{3} \zeta(3) \ln(2)
 \right. \nonumber \\ & & \hspace*{1cm} \left.
   +n_f \left[ \frac{55}{20736} \pi^5 -\frac{1}{3456} \pi^3
   +\frac{35}{27} \zeta(3)-\frac{28}{9} \zeta(3) \ln (2) \right] \right)
\nu^2 \nonumber \\ & & \hspace*{-70pt}
  +\left(-h_4- \frac{933}{5120} \pi^7 +\frac{5}{768} \pi^5 +\frac{1}{128} \pi^3
  -1905 \zeta(7) + \zeta(5) \left(
\frac{2077}{18}  -\frac{31}{6}  \pi^2
 -\frac{682}{3} \ln(2) \right) -\frac{539}{6} \zeta(3)^2
  \right.
\nonumber \\ & &  \hspace*{-2cm}\left.
 +n_f \left[ -\frac{671}{414720} \pi^7 + \frac{5}{20736}\pi^5+\frac{1}{3456}
 \pi^3- \zeta(5) \left(  \frac{155}{27}  -\frac{124}{9} \ln (2) \right)
 +\frac{49}{9}\zeta(3)^2 \right] \right) \nu^4 \ + \ \cdots ,
 \label{eq7} \end{eqnarray} 
where
\begin{equation}
h(\nu)\ =\ 2 \int_0^1 \frac{dx}{\sqrt{x}(1+x)} \cos(\nu\ln x)
 {\mathrm Li}_2(1-x) 
 \ = \ h_0 +h_2 \nu^2 + h_4 \nu^4 \ + \ \cdots , \end{equation}
${\mathrm Li}_2$ being the dilogarithm function.

\begin{figure}
\begin{center}
\leavevmode
\hbox{\epsfxsize=4.8 in
\epsfbox{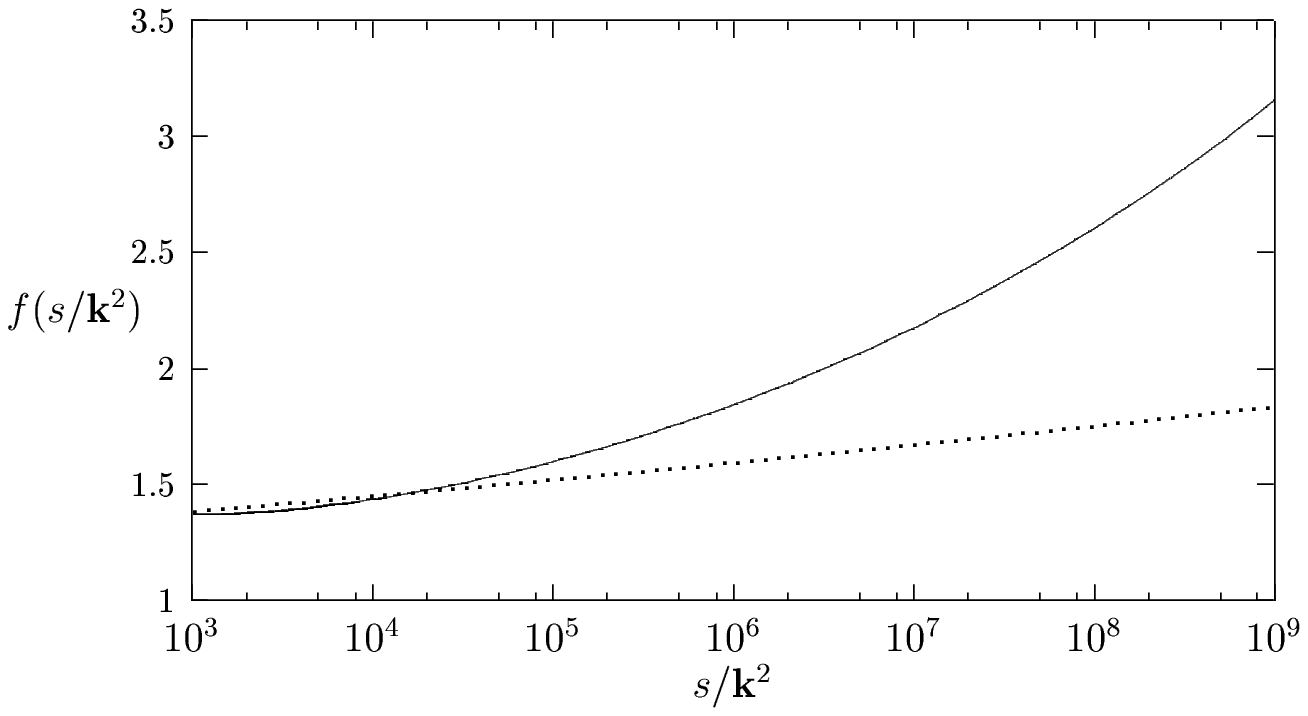}}
\caption{}
\end{center}
\end{figure}

\noindent Numerical integration gives
$$ h_0=4.115, \ \ h_2=-24.92, \  \ h_4= 87.47, $$
so that if we write
\begin{equation}
\omega(\nu) \ = \omega_0  \ +  \ a \nu^2  \ -  \ b \nu^4  \ +  \ \cdots,
\end{equation}
 then for $\alpha_s=0.15$ and three flavours we obtain
$$ \omega_0=0.021, \ \ a=4.19, \ \ b=47.4 $$
The solution to Eq.(\ref{eq1}) up to ${\cal O}(1/\ln^2 s)$ is then
obtained by expanding around the saddle point at $\nu^2=a/2b$ and gives
\begin{equation}
f(s,{\mathbf k,  k^\prime }) \sim 
\frac{1}{\sqrt{{\mathbf k}^2 {\mathbf k^\prime}^2}}
s^{(\omega_0+a^2/4b)} \frac{1}{\sqrt{ a \ln s}}
\exp(\frac{3b}{4a^2\ln s})
 \cos\left( \sqrt{\frac{a}{2b}}\left(1-\frac{3b}{4a^2\ln s}\right)\ln r\right),
\label{eq10} \end{equation}
where
$$r \ = \ \frac{\sqrt{\alpha_s({\mathbf k^\prime}^2)} {\mathbf k}^2}
   {\sqrt{\alpha_s({\mathbf k}^2)} {\mathbf k^\prime}^2}. $$
We can see immediately from Eq.(\ref{eq10}) that the power behaviour
of $s$ has an exponent of $0.12$ for $\alpha_s=0.15$ rather than
 $0.02$ which one obtains from a consideration of $\omega(0)$.
These two behaviours are plotted for against $s$ for the case
 ${\mathbf k}^2={\mathbf k^\prime}^2$. $s$ is normalized 
by a typical square transverse momentum  ${\mathbf k}^2$
 ( in deep inelastic scattering $1/x$ plays the role of $s/\mathbf{k}^2$).
The solid line is the plot of Eq.(\ref{eq10}), whereas the dotted line
is the $\omega_0$ only behaviour. These have been
normalized so that they are equal at $s/{\mathbf k}^2=10^3$.
We note that at  $s/{\mathbf k}^2=10^5$, which represents
the smallest $x$ values that one might currently expect at HERA,
  the difference between the two are not very significant. However the
two curves are significantly different if we go up to much larger energies
corresponding to rapidity gaps which one might hope to achieve in
future hadron colliders such as LHC.

\begin{figure}
\begin{center}
\leavevmode
\hbox{\epsfxsize=4.8 in
\epsfbox{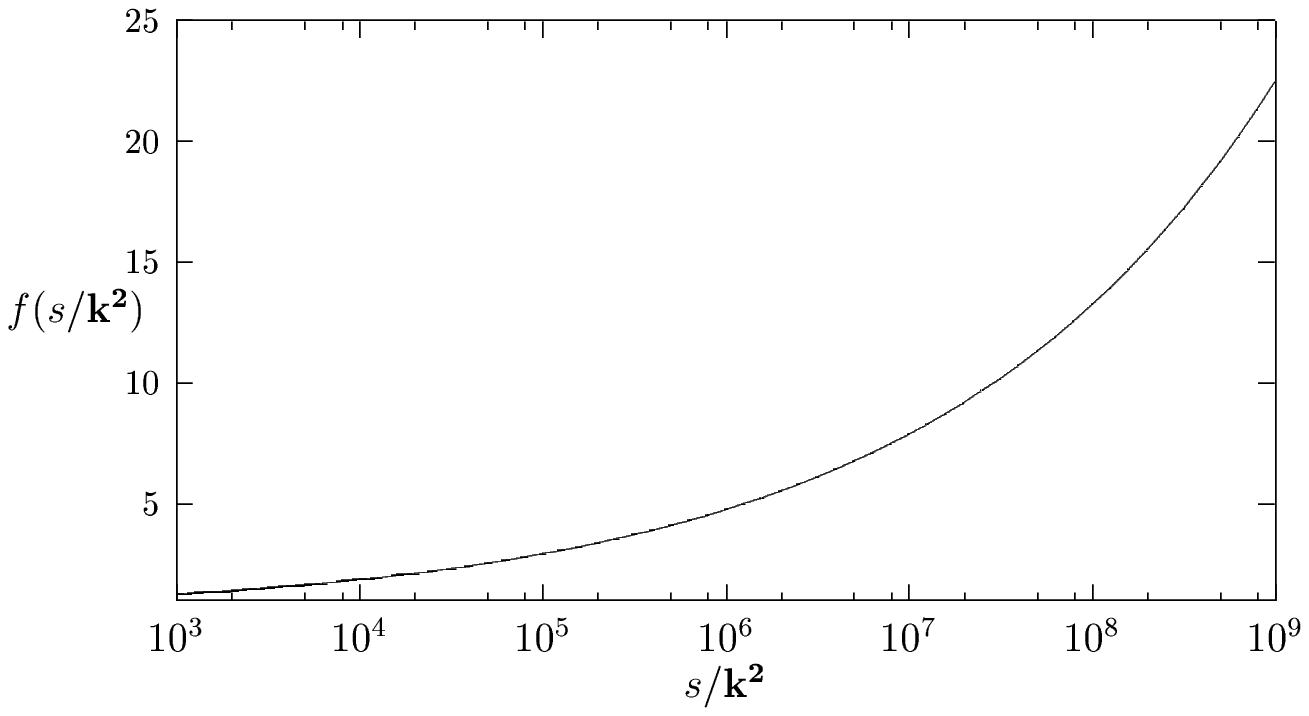}}
\caption{}
\end{center}
\end{figure}

The exponent of $0.1$ for this value of $\alpha_s$ is slightly below
the value of $\omega_{max}$ as can be seen from Fig. 2, and the value
of $\nu$ at which this maximum value occurs is a little higher than
the position of the saddle point,  $\nu_{saddle}=0.21$. For larger values
of the coupling one does not expect this saddle point approximation to 
be reliable. Here one expects to sample larger values of $\nu$ for
which the quartic approximation to $\omega$ begins to break down.
Rather we use Fig. 2 to read off the the corresponding $\omega_{max}$
and allow for the fact that the dominant value of $\omega$ is expected
to be slightly below this value. Thus for $\alpha_s \approx 0.22$ we
expect an exponent of 0.2. 
This is indeed what is found by a direct numerical solution of Eq.(\ref{eq1})
with  $\alpha_s$ taken to be 0.22, shown in Fig. 4.
This corresponds   to a pomeron intercept
of 1.2, which is consistent with the value quoted by the H1 collaboration
\cite{H1} from studies of the diffractive proton structure function
\footnote{On the other hand the Zeus collaboration \cite{zeus}
 favours a smaller value of the intercept consistent with the soft pomeron.}
at $Q^2=18 \ \mathrm{GeV}^2$ (for which $\alpha_s \approx 0.22$).

To summarize we have shown that although the order $\alpha_s^2$
correction to the eigenvalues of the BFKL kernel do indeed significantly
suppress the growth of the diffractive amplitude with increasing $s$,
this suppression is much less  than that which is obtain from a
consideration of the $\nu=0$ components. For $\alpha_s \approx 0.22$
we predict a pomeron intercept of 1.2, which is consistent with
experimental data on diffractive deep-inelastic scattering.

\end{document}